\documentclass[conference]{IEEEtran}
\IEEEoverridecommandlockouts
\usepackage{cite}
\usepackage{amsmath,amssymb,amsfonts}
\usepackage{algorithmic}
\usepackage{graphicx}
\usepackage{textcomp}
\usepackage{xcolor}
\def\BibTeX{{\rm B\kern-.05em{\sc i\kern-.025em b}\kern-.08em
    T\kern-.1667em\lower.7ex\hbox{E}\kern-.125emX}}

\usepackage{makecell}
\usepackage{array}
\usepackage{subfigure}
\usepackage{amsmath}
\usepackage{color}
\usepackage{mathtools}

\usepackage[ruled,linesnumbered]{algorithm2e}

\begin{document}

\title{Signed Graph Convolutional Network}

\author{\IEEEauthorblockN{Tyler Derr}
\IEEEauthorblockA{\textit{Data Science and Engineering Lab} \\
\textit{Michigan State University}\\
derrtyle@msu.edu}
\and
\IEEEauthorblockN{Yao Ma}
\IEEEauthorblockA{\textit{Data Science and Engineering Lab} \\
\textit{Michigan State University}\\
mayao4@msu.edu}
\and
\IEEEauthorblockN{Jiliang Tang}
\IEEEauthorblockA{\textit{Data Science and Engineering Lab} \\
\textit{Michigan State University}\\
tangjili@msu.edu}
}
\maketitle

\begin{abstract}
Due to the fact much of today's data can be represented as graphs, there has been a demand for generalizing neural network models for graph data. One recent direction that has shown fruitful results, and therefore growing interest, is the usage of graph convolutional neural networks (GCNs). They have been shown to provide a significant improvement on a wide range of tasks in network analysis, one of which being node representation learning. The task of learning low-dimensional node representations has shown to increase performance on a plethora of other tasks from link prediction and node classification, to community detection and visualization. Simultaneously, signed networks (or graphs having both positive and negative links) have become ubiquitous with the growing popularity of social media. However, since previous GCN models have primarily focused on unsigned networks (or graphs consisting of only positive links), it is unclear how they could be applied to signed networks due to the challenges presented by negative links. The primary challenges are based on negative links having not only a different semantic meaning as compared to positive links, but their principles are inherently different and they form complex relations with positive links. Therefore we propose a dedicated and principled effort that utilizes balance theory to correctly aggregate and propagate the information across layers of a signed GCN model. We perform empirical experiments comparing our proposed signed GCN against state-of-the-art baselines for learning node representations in signed networks. More specifically, our experiments are performed on four real-world datasets for the classical link sign prediction problem that is commonly used as the benchmark for signed network embeddings algorithms. 
\end{abstract}


\section{Introduction}\label{sec:introduction}

Recently there has been a large and growing interest of generalizing neural network models to structured data, with one of the most prevalent structures being graphs (such as those found in social media). The idea of generalizing neural network models to graph structures has lately started to become more developed by overcoming the difficulties and trade-offs previously associated with fast heuristics compared to slow and more principled approaches. Graph convolutional neural networks (GCNs)~\cite{Niepert-etal2016,Defferrard-etal2016,Kipf-Welling2016,Duvenaud-etal2015,Hamilton-etal2017,Bruna-etal2014} are modeled after the classical convolutional neural networks~\cite{Lecun-etal1998}. The first GCN introduced for learning representations at the node level was in~\cite{Kipf-Welling2016}, where they utilized GCNs for the semi-supervised node classification problem. Furthermore, learning low-dimensional node representations have been previously proven to be useful in many network analysis tasks beyond node classification~\cite{Wang-etal2016a,Perozzi-etal2014}, such as link prediction~\cite{Wang-etal2016,Grover-Leskovec2016}, community detection\cite{Cao-etal2015,Wang-etal2017a}, and visualization~\cite{Kunegis-etal2010,Wang-etal2016}. 

Previous work has mostly focused on using GCNs for unsigned graphs (or graphs consisting of only positive links). However, especially with the ever growing popularity of online social media, signed graphs are becoming increasingly ubiquitous. This naturally leads the question as to whether unsigned GCNs are suitable to be used on signed networks. Unfortunately, there are many reasons as to why unsigned GCNs are not capable of learning meaningful node representations in signed networks. First, it is unclear how they would handle the availability of negative links in signed networks, and furthermore, negative links invalidate some of the underlying key assumptions of GCNs. For example, GCNs designed for unsigned networks learn a node representation using the fundamental social theory homophily~\cite{Mcpherson-etal2001}, which states users having connections are more likely to be similar than those without links. Hence, the aggregation processes of GCNs use local neighborhood information when constructing the low-dimensional embedding for each node. However, homophily may not be applicable to signed networks~\cite{Tang-etal2014}. Instead, in signed networks, there are specific social theories and principles defined in the context of having both positive and negative links. Therefore dedicated efforts are needed for redesigning GCNs specifically for signed networks.

Although it is now clear that GCNs will need to be specifically redesigned to provide the same fruitful performance as previously shown in unsigned networks when applied to signed networks, there are still tremendous challenges to overcome. When designing signed GCNs the primary challenges are: (1) how to correctly handle negative links, since their properties are inherently different than those of positive links; and (2) how to combine the positive and negative links into a single coherent model to learn effective node representations. Thus, we turn our attention towards social theories specific to signed networks (similarly to how the unsigned models were constructed using unsigned theories like homophily). More specifically, one fundamental signed network social theory that had been developed in social psychology is balance theory~\cite{Heider-1946,Cartwright-Haray1956}. If we can harness the power of this signed network social theory, which provides a better understanding of negative links and how they form complex relations with positive links, then this offers the opportunity to solve these two challenges when incorporating these ideas into our framework of designing a signed GCN. Our major contributions are listed as follows:
\begin{itemize}
\item Propose a Signed Graph Convolutional Network (SGCN) that is constructed based on balance theory to correctly integrate negative links during the aggregation process;
\item Construct an objective function for our SGCN based on signed network social theories to easily learn an effective low-dimensional representation for each node in the network;
\item Conduct experiments on four real-world signed networks to comprehensively demonstrate the effectiveness of our proposed SGCN framework.
\end{itemize}

The rest of the paper is organized as follows. In Section~\ref{sec:problemstatement}, we define the problem of signed network embedding and introduce the notations used in the remainder of the paper. We present our proposed signed GCN framework in Section~\ref{sec:framework}, which consists of giving a brief overview on unsigned GCNs, how to simultaneously capture and combine both the positive and negative links during the aggregation process, and finally introduce our Signed Graph Convolutional Network (SGCN) along with a discussion on how to optimize its parameters. In Section~\ref{sec:experiments}, experiments are performed to empirically evaluate the effectiveness of our framework for learning node embeddings. We discuss related work on signed network embedding and graph convolutional networks in Section~\ref{sec:relatedwork}. Finally, we conclude and discuss future work in Section~\ref{sec:conclusion}.

\section{Problem Statement}\label{sec:problemstatement}
Let $\mathcal{G}=(\mathcal{U},\mathcal{E}^+, \mathcal{E}^-)$ be a signed network, where $\mathcal{U} = \{u_1, u_2, \dots u_n \}$ represents the set of $n$ nodes while $\mathcal{E}^+ \subset \mathcal{U} \times\mathcal{U}$ and $\mathcal{E}^- \subset \mathcal{U} \times \mathcal{U}$ denote the sets of positive and negative links, respectively. Note that $\mathcal{E}^+ \cap \mathcal{E}^- = \emptyset$, in other words, a pair of nodes cannot have both positive and negative links simultianeously. We use $\mathbf{A} \in \mathbb{R}^{n \times n}$ to denote the adjacency matrix of the signed network $\mathcal{G}$, where $\mathbf{A}_{ij} = 1$ means there exists a positive link from $u_i$ to $u_j$, $\mathbf{A}_{ij} = -1$ denotes a negative link, and $\mathbf{A}_{ij} = 0$ otherwise (meaning no link from $u_i$ to $u_j$). In Table~\ref{tab:notations} we further summarize the major notations used throughout this work. 

With the aforementioned notations and definitions, we can now formally define the problem of signed network embedding as follows:\\

\noindent \textit{Given a signed network $\mathcal{G}=(\mathcal{U},\mathcal{E}^+, \mathcal{E}^-)$ represented as an adjacency matrix $\mathbf{A} \in \mathbb{R}^{n \times n}$, we seek to discover a low-dimensional vector for each node as }
\begin{align}
F: \mathbf{A} \rightarrow \mathbf{Z}
\end{align}
where $F$ is a learned transformation function that maps the signed network's adjacency matrix $\mathbf{A}$ to a $d$-dimensional representation $\mathbf{Z} \in \mathbb{R}^{n \times d}$ for the $n$ nodes of the signed network.

\begin{table}
\begin{center}
\caption{Notations.}
\vspace{-0.001in}
\label{tab:notations}
\begin{tabular}{|l|l|}
	\hline
	Notations & Descriptions\\
	\hline
	$\mathbf{A}$ & Adjacency matrix\\
    $\mathbf{Z}$ & Low-dimensional representation of signed network $\mathcal{G}$ \\
    $B_i(l)~(U_i(l))$ & The set of users that can be reached from $u_i$ \\
    &along a (un)balanced path of length $l$. \\
    $B(l)~(U(l))$ & The aggregator responsible for incorporating \\
    &the information from the set of users $B_i(l) (U_i(l))$\\
	$\mathbf{z}_i$ & The final embedding of user $u_i$ \\
    $\mathcal{N}^+_i ~(\mathcal{N}^-_i)$ & Set of positive (negative) neighbors of $u_i$ \\
    $\mathbf{h}^{B(l)}_i ~(\mathbf{h}^{U(l)}_i)$ & The (un)balanced representation of $u_i$ at the $l^\text{th}$ layer \\
    $\mathbf{W}^{B(l)} ~(\mathbf{W}^{U(l)})$ & Weight matrices used for learning how to propagate \\
    & (un)balanced information in the $l^{\text{th}}$ layer \\
	\hline
\end{tabular}
\vspace{-0.1in}
\end{center}
\end{table}
\section{The Proposed Framework}\label{sec:framework}

Graph convolutional neural networks have recently started to become more developed and have already shown their superiority in extracting and aggregating information from graph data. Their use cases spread over the vast field of network analysis, but one such domain that has shown to be very influential recently is network embedding. The discovery of representative low-dimensional features for each node in the network has previously shown to enhance many tasks from link prediction and node classification, to community detection and visualization. However, previous work has mostly focused on constructing GCNs for unsigned networks. Due to the inherent differences between unsigned and signed networks, this leaves a gap that we seek to bridge with the development of a signed graph convolutional network (SGCN). 

Even with dedicated efforts towards the construction of a GCN specific to signed networks, there are still tremendous challenges we must face and overcome. The first of which is figuring out how we can correctly incorporate negative links during the aggregation process. We cannot simply treat the negative links the same as positive links, since their properties and semantic meaning vastly differ. The second challenge is how we can combine the two sets of links (i.e., positive and negative) into a single coherent model. This combination is essential because certainly positive and negative links interact in the network structure in complex ways and indeed are not segregated and isolated from each other. 

In this work we propose to go to the roots of signed network analysis and utilize one of the most fundamental and indispensable signed social theories developed in social psychology, balance theory~\cite{Cartwright-Haray1956,Heider-1946}. We harness balance theory to construct a bridge to connect the gap between the ongoing development of GCNs for unsigned networks and signed networks. In the remainder of this section we will first briefly discuss a general GCN framework in the unsigned network setting and discuss the relationships of this framework to the structure of signed networks. Then we introduce balance theory and how we can use this signed social theory to correctly capture both positive and negative links simultaneously during the aggregation process. Thereafter, we present how to learn the parameters of our SGCN -- first through the construction of an objective function designed to effectively learn the node representations in signed networks, and finally discussing the optimization procedure taken to optimize our proposed objective. 


\subsection{Unsigned Graph Convolutional Networks}

Currently, most GCNs have a similar structure in that they utilize a convolutional operator that can share weights across all locations in the graph. The benefits of this neural network structure in graphs as compared to the cumbersome fully connected models are at least three fold: 1) it avoids the parameter explosion associated with fully connected layers (especially when handling larger graphs); 2) it allows for parameter sharing across the network to avoid overfitting; and 3) a single GCN is capable of handling as input graphs of varying structures and even sizes (in terms of the number of nodes and edges).

Typically the architecture of an unsigned GCN for learning node representations is of the form shown in Algorithm~\ref{alg:unsignedGCN}. 
In the process of generating the $d^{out}$-dimensional embedding matrix ${\bf Z} \in \mathbb{R}^{x \times d^{out}}$, they make use of the unsigned adjacency matrix ${\bf A} \in \mathbb{R}^{n \times n}$ and a feature matrix ${\bf X} \in \mathbb{R}^{n \times d^{in}}$, where $d^{in}$ is the length of feature vector ${\bf x}_i$ for user $u_i$. The matrices ${\bf H}^{(l)} \in \mathbb{R}^{n \times d^{out}}$ for $l \in \{1, \dots, L\}$ represent the hidden representations for each of the $n$ nodes of the graph at each layer $l$ of the GCN. On line 1 we set the initial representation ${\bf H}^{(0)}$ equal to ${\bf X}$ to ease the notations in the remainder of the algorithm. Then, on line 2 we loop updating the parameters of the GCN until convergent. Inside this loop, for each update iteration we propagate the graph features through the $L$ layers of the GCN using the unsigned adjacency matrix ${\bf A}$ and neighborhood aggregation function $f()$. Note that the function $f()$ is where the variations of GCNs primarily differ. 
Finally, after the model converges, the embedding is taken as the last layer's representation matrix ${\bf H}^{(L)}$.

{\bf Limitations of unsigned GCN for signed networks:} Given the above discussion on the unsigned GCN framework, we note that in relation to signed networks this would be similar to applying the unsigned GCN on the positive only adjacency matrix ${\bf A}^+$ where ${\bf A}^+_{ij} = 1$ if there exists a positive link between users $u_i$ and $u_j$, and 0 otherwise (i.e., when there exists either a negative link or no link between them). However, this would ignore the negative links. 

Initially, our thoughts may lead to some na\"ive approaches of handling the negative links by either ignoring them, treating them the same or the negation of positive links, or separately applying the GCN framework to first the positive network, and then the negative network with finally combining them at the end stage. However, each of these methods is either based on incorrect assumptions or ignoring parts of the rich information awaiting to be extracted from the complex network structure of signed networks towards the learning of an advantageous low-dimensional representation. For example, trivially treating the negative links the same as the positive links would be an incorrect assumption, since negative links have been shown to have different principles and semantically represent vastly different meanings. Similarly, treating negative links as the negation of positive links is likely an incorrect assumption~\cite{Tang-etal2014}. This leaves the last two initial thoughts of ignoring the negative links or applying an unsigned GCN separately on the positive only and negative only networks, but intuitively the first choice is certainly ignoring a large amount of information, and based on signed social theories~\cite{Leskovec-etal2010a,Heider-1946,Cartwright-Haray1956}, there exist complex relations between the positive and negative links that if extracted, can provide fruitful results~\cite{Leskovec-etal2010,Kunegis-etal2009}. Therefore, next we will discuss one such signed social theory, balance theory~\cite{Heider-1946,Cartwright-Haray1956} and how we propose to harness it for capturing both the positive and negative links coherently together during the aggregation process.  

 \begin{algorithm}[t]
 \small
 	\DontPrintSemicolon
 	\KwIn{An unsigned network adjacency matrix ${\bf A} \in \mathbb{R}^{n \times n}$; a feature matrix ${\bf X} \in \mathbb{R}^{n \times d^{in}}$; number of aggregation layers $L$; neighborhood aggregation function $f()$}
 	\KwOut{Low-dimensional representation matrix ${\bf Z} \in \mathbb{R}^{n \times d^{out}}$}
 	${\bf H}^{(0)} \leftarrow {\bf X}$\;
    \While{not convergent}
    {

		\For{$l \in \{0, \dots ,L-1\}$}
    	{
    		${\bf H}^{(l+1)} \leftarrow f({\bf H}^{(l)},{\bf A})$\;
    	}
	
    	Update GCN parameters based on $loss({\bf H}^{(L)})$
	}
    ${\bf Z} \leftarrow {\bf H}^{(L)}$\;
  \caption{Typical Unsigned GCN Framework.}
 \label{alg:unsignedGCN}
 \end{algorithm}

\subsection{Aggregation paths with positive and negative links}
Balance theory dates back to the early seminal work in~\cite{Heider-1946} and later generalized in~\cite{Cartwright-Haray1956} having a graph theoretical foundation. 
In general, balance theory implies ``the friend of my friend is my friend'' and ``the enemy of my friend is my enemy''. The theory classifies cycles in a signed network as being either \textit{balanced} or \textit{unbalanced}, where a balanced cycle consist of an even number of negative links while a cycle having an odd number of negative links is considered unbalanced. We first introduce the four possible cycles that can be formed in a signed network and followed by the introduction of our definition of balanced and unbalanced paths inspired by the general definition of balanced and unbalanced cycles.

\begin{figure}
	\begin{center}
		\includegraphics[scale=0.4]{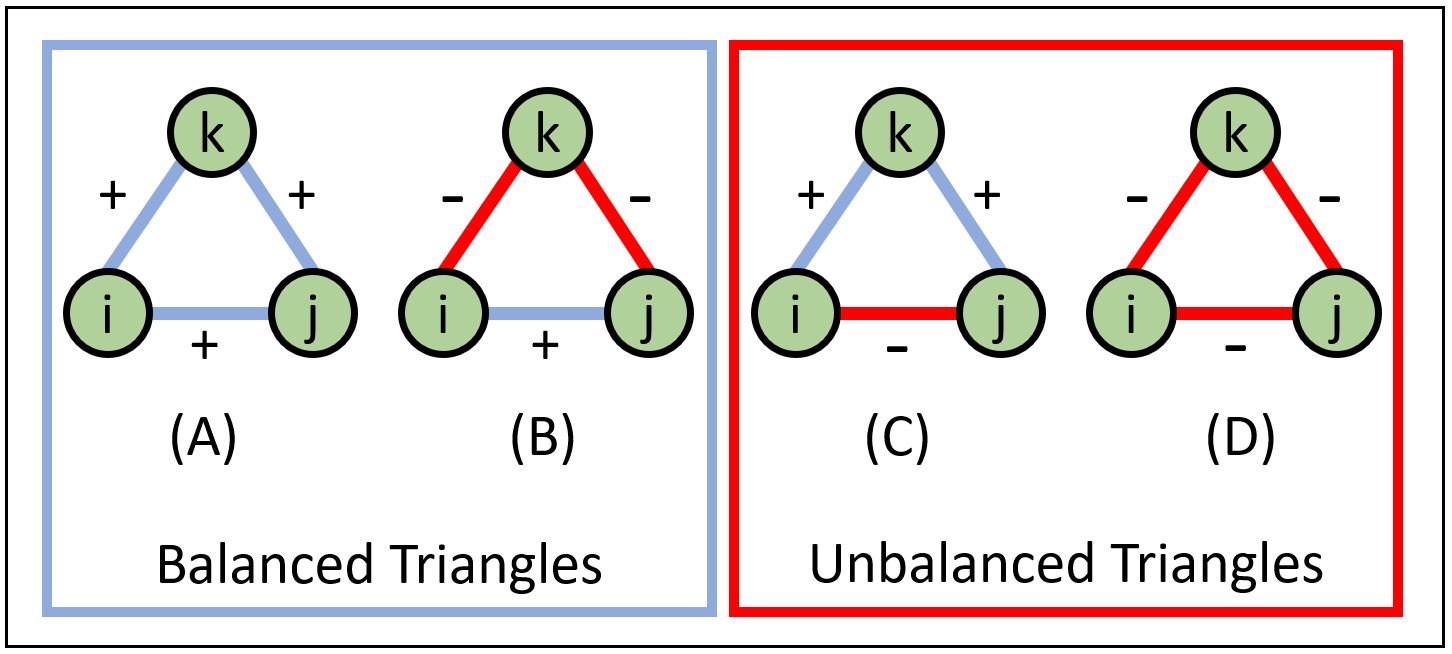}
	\end{center}
	\caption{The four undirected signed triangles types according to balance theory.}
	\label{fig:balancetheory}
\end{figure}

\begin{figure}
	\begin{center}
		\includegraphics[scale=0.45]{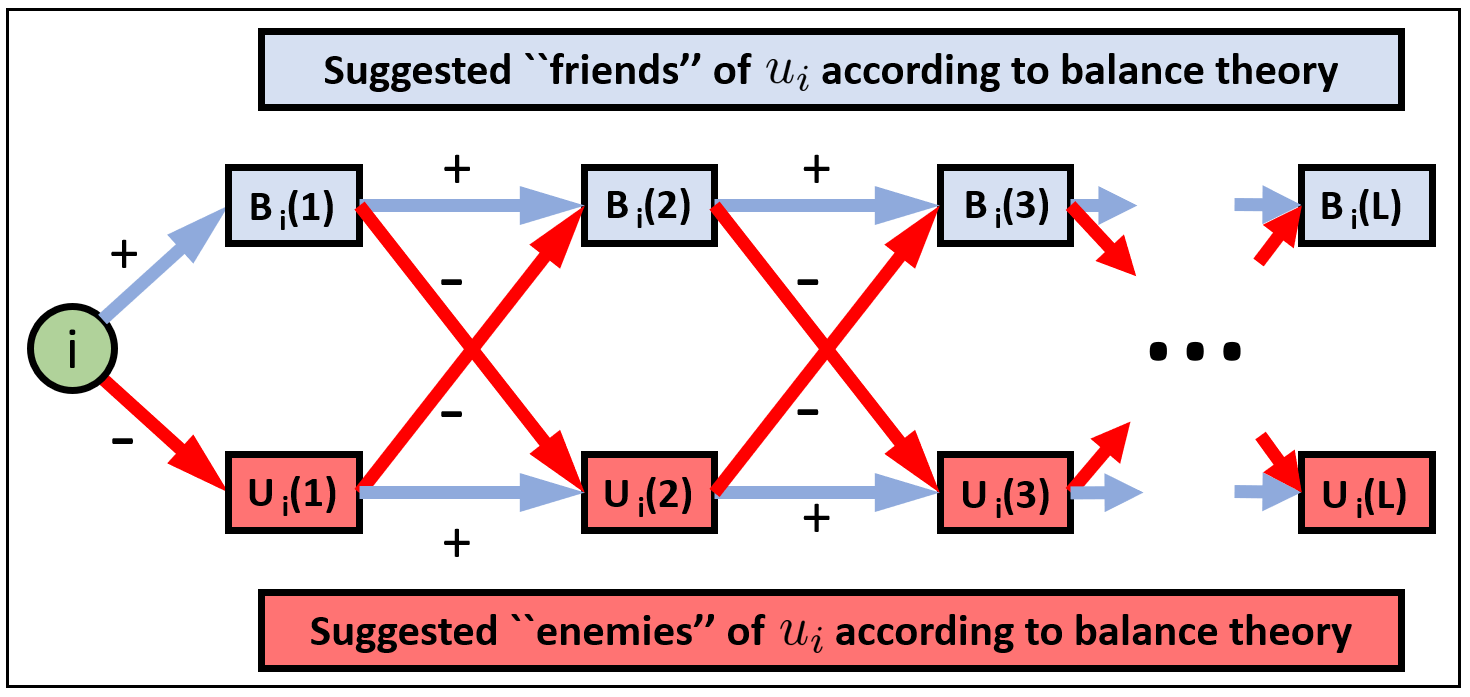}
	\end{center}
	\caption{An illustration of the link structure that leads to users being a part of the different aggregations based on balance and unbalanced paths of different lengths. }
	\label{fig:bal_unbal_paths}
\end{figure}

In Figure~\ref{fig:balancetheory} we can see that triangles (A) and (B) are balanced, while (C) and (D) are unbalanced. We propose to denote a \textit{balanced path} as one that consists of an even number of negative links, and similarly an \textit{unbalanced path} being one that has an odd number of negative links. With these definitions, along with balance theory, we can see that if we had a path of length $l$ from $u_i$ to $u_j$ that had an even number of negative links, then balance theory would suggest a positive 
link between $u_i$ an $u_j$. An example of a balanced path can be seen in Figure~\ref{fig:balancetheory} triangle (B), where the path of length two from $u_i$ to $u_j$ (through $u_k$) consists of two negative links and thus balance theory would suggest a positive link connecting $u_i$ and $u_j$ (to result in a balanced cycle between users $u_i$, $u_k$, and $u_j$). From the context of user $u_i$ we would then place $u_j$ into the set $B_i(2)$, which we use to denote the set of users that can be reached from user $u_i$ along a balanced path of length 2. In the general case, users that can be reached from $u_i$ along a balanced (or unbalanced) path of length $l$ we place in the set $B_i(l)$ (or $U_i(l)$). In Figure~\ref{fig:bal_unbal_paths} we provide an illustration of how all the signed paths of a given length would place users along paths from $u_i$ into their respective sets. Note that the arrows are only used to aid the illustration and that our definition is based on the more general undirected setting.

Before continuing, let us define $\mathcal{N}^+_i$ to be the set of positive neighbors of a user $u_i$, i.e., $u_j \in \mathcal{N}^+_i$ if ${\bf A}_{ij} = 1$. We similarly denote the set of negative neighbors for user $u_i$ as $\mathcal{N}^-_i$, where $u_j \in \mathcal{N}^-_i$ when ${\bf A}_{ij} = -1$. In Figure~\ref{fig:bal_unbal_paths} we can see that when having a balanced path of length $l$ from $u_i$ to some user $u_k$ (i.e., $u_k \in B_i(l)$), then all the positively linked neighbors of $u_k$ (which we denoted as the set $\mathcal{N}^+_k$) would be placed in $B_i(l+1)$. This is because adding a positive link to a balanced path (i.e., a path consisting of an even number of negative links) still results in a balanced path, but just of additional length. Similarly when adding a negative link to a balanced path, we obtain an unbalanced path.

Another key observation from Figure~\ref{fig:bal_unbal_paths} is how we can obtain the balanced and unbalanced sets $B_i(l+1)$ and $U_i(l+1)$ of length $l+1$, respectively for user $u_i$, from the sets $B_i(l)$ and $U_i(l)$ of length $l$. Below we provide a recursive definition for calculating the balanced and unbalanced sets from the perspective of user $u_i$ as follows:\\

When $l = 1$
\begin{align}
&B_i(1) = \{u_j ~|~ u_j \in \mathcal{N}^+_i\} \nonumber \\ 
&U_i(1) = \{u_j ~|~ u_j \in \mathcal{N}^-_i\} \nonumber
\end{align}
For $l > 1$
\begin{align}\label{eq:bal_unbal_sets}
B_i(l+1) =   ~& \{u_j ~|~ u_k \in B_i(l) ~\text{and}~ u_j \in \mathcal{N}^+_k\} \nonumber \\ 
		\cup ~& \{u_j ~|~ u_k \in U_i(l) ~\text{and}~ u_j \in \mathcal{N}^-_k\} \nonumber \\
\nonumber\\
U_i(l+1) =   ~& \{u_j ~|~ u_k \in U_i(l) ~\text{and}~ u_j \in \mathcal{N}^+_k\} \nonumber \\
		\cup ~& \{u_j ~|~ u_k \in B_i(l) ~\text{and}~ u_j \in \mathcal{N}^-_k\} \nonumber \\
\end{align}

Given the above definition, we again note that the users in the balanced sets (which are reached along balanced paths) for a user $u_i$ are those that either: 1) have a positive link directly to $u_i$; or 2) those that balance theory would suggest a 
positive link between them since they have an even number of negative links along the path connecting them. For the unbalanced sets the definition is similar, except with direct/suggested negative links. We note that these definitions, based upon balance theory, now allow us a principled way of aggregating and propagating information in signed networks using balanced and unbalanced paths/sets. Next we will propose aggregation functions for our signed GCN and follow with the rest of the details of our framework.

\subsection{Signed Graph Convolutional Network}
Before formalizing our signed graph convolutional network, we provide some insights and intuitions behind the construction in light of balanced and unbalanced sets and paths. The first insight is that in unsigned GCNs, when constructing a node representation, they aggregate their immediate local neighbors' information into a single representation and then through the use of multiple layers, propagate this in the network allowing a node to incorporate information from a multi-hop neighborhood (where the number layers in the GCN denotes the number of hops away information is being aggregated from). However, in signed networks, we cannot categorize all users the same. This is because semantically users that are connected through positive links to $u_i$ are thought of as their ``friends'' while neighbors across negative links are their ``enemies''. Similarly, for users in $u_i$'s balanced sets, balance theory would suggest they are their ``friends'' (even though they are not directly linked) and those in $u_i$'s unbalanced sets are suggested to be their ``enemies'' based on this social theory. This phenomenon can be visualized in Figure~\ref{fig:bal_unbal_paths}. Therefore, we propose rather than maintaining a single representation for each node, we keep a representation of both their ``friends'' and ``enemies'', which successfully incorporates both the positive and negative links and gives a more thorough representation of a given user. 

\begin{figure}
	\begin{center}
		\includegraphics[scale=0.43]{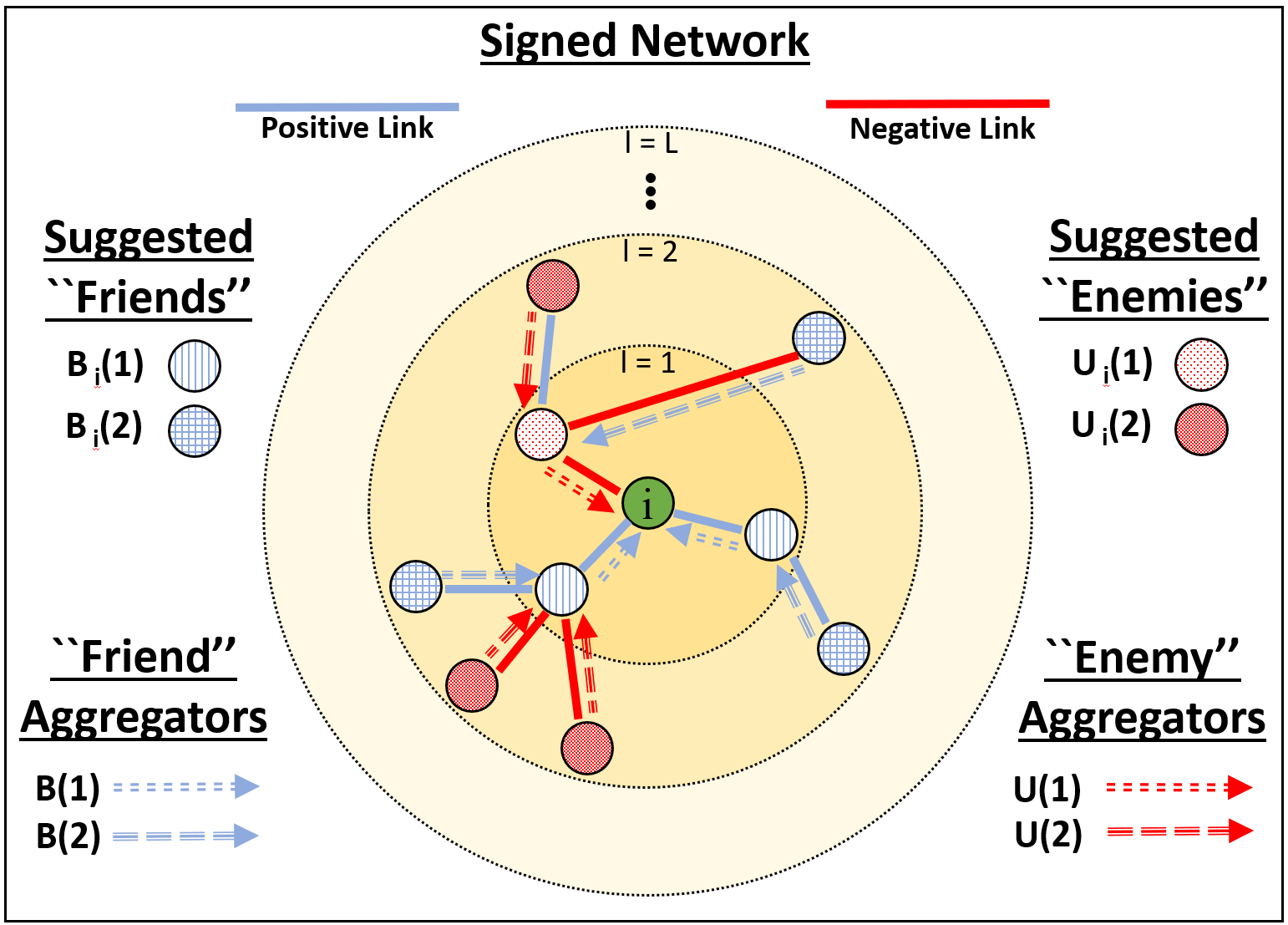}
	\end{center}
	\caption{An illustration of how SGCN aggregates neighbor information in a signed network. }
	\label{fig:sgcn_aggregation}
\end{figure}

In Figure~\ref{fig:sgcn_aggregation} we provide an illustration of how we plan to aggregate and propagate information in a signed networks. Note that the circles labeled $l=1,2,\dots,L$ are used to denote how many hops away the user is from $u_i$ and simultaneously denotes at which layer in our signed GCN that user's information will be incorporated into the two learned representations for user $u_i$. We can observe that we could have a separate aggregator responsible for incorporating the information from each respective balanced and unbalanced sets. For example, in the first layer of Figure~\ref{fig:sgcn_aggregation} we can see that the two positive neighbors of $u_i$ will be incorporated into the level one ``friend'' representation through the use of aggregator $B(1)$. Similarly $u_i$'s single negatively linked neighbor is used for learning the level one ``enemy'' representation. Then, through the use of a second layer in our GCN, we can incorporate the two-hop neighbors. However, the crucial step here is that we must aggregate the information of these neighbors correctly to adhere to balance theory according to our defined balanced and unbalanced paths/sets. Therefore we employ a second set of aggregators, namely $B(2)$ and $U(2)$ which will help propagate the information from users in sets $B_i(2)$ and $U_i(2)$, respectively. Notice that just as shown in Figure~\ref{fig:bal_unbal_paths} users being included by the $B(2)$ aggregator are the users who are along a path of two consecutive positive links, or two consecutive negative links, because they are both suggested as ``friends'' according to balance theory. On the other hand, aggregator $U(2)$ (which is gathering information from users in the set $U_i(2)$) seeks to utilize the information from users along paths that consist of one positive and one negative link (in either ordering, since both fall into set $U_i(2)$). Now we can more formally discuss the aggregation functions used by our proposed SGCN.

While aggregating and propagating information in our SGCN, we will maintain two representations at each layer, one for the corresponding balanced set of users (i.e., suggested ``friends''), and one for the users in the respective unbalanced set (i.e., suggested ``enemies''). Similar to the unsigned GCN, we use ${\bf h}^{(0)}_i \in \mathbb{R}^{d^{in}}$ to represent the initial $d^{in}$ node features for user $u_i$. Thus, for the first aggregation layer (i.e, when $l=1$), we utilize the following:

\begin{align}
{\bf h}_i^{B(1)} = \sigma \Bigg( {\bf W}^{B(1)} \Big[ \sum\limits_{j \in \mathcal{N}_i^+ } \frac{{\bf h}_j^{(0)}}{|\mathcal{N}_i^+ |}, {\bf h}_i^{(0)} \Big]  \Bigg)
\end{align}
\begin{align}
{\bf h}_i^{U(1)} = \sigma \Bigg( {\bf W}^{U(1)} \Big[ \sum\limits_{k \in \mathcal{N}_i^-} \frac{{\bf h}_k^{(0)}}{|\mathcal{N}_i^-|}, {\bf h}_i^{(0)} \Big]  \Bigg)
\end{align}

where $\sigma()$ is a non-linear activation function, ${\bf W}^{B(1)}, {\bf W}^{U(1)} \in \mathbb{R}^{d^{out} \times 2d^{in}}$ are the linear transformation matrices responsible for the ``friends'' and ``enemies'' coming from sets $B_i(1)$ and $U_i(1)$, respectively, and $d^{out}$ is the length of the two internal hidden representations. More specifically, for determining the hidden representation ${\bf h}^{B(1)}_i$ we also concatenate the hidden representation of user $u_i$ (i.e., ${\bf h}^{(0)}_i$) along with the mean of the users in set $B_i(1)$. In all subsequent layers, the aggregation is more complex, just as the definition of $B_i(l)$ an $U_i(l)$ were more complex when $l>1$ in Eq.~(\ref{eq:bal_unbal_sets}). This is similarly due to the cross linking of negative links as seen in Figure~\ref{fig:bal_unbal_paths}. The aggregations for $l>1$ are defined as follows:

\begin{align}
{\bf h}_i^{B(l)} = \sigma \Bigg( {\bf W}^{B(l)} \Big[ \sum\limits_{j \in \mathcal{N}_i^+} \frac{{\bf h}_j^{B(l-1)}}{|\mathcal{N}_i^+|}, \sum\limits_{k \in \mathcal{N}_i^-} \frac{{\bf h}_k^{U(l-1)}}{|\mathcal{N}_i^-|}, {\bf h}_i^{B(l-1)}\Big] \Bigg) 
\end{align}

\begin{align}
{\bf h}_i^{U(l)} = \sigma \Bigg( {\bf W}^{U(l)} \Big[ \sum\limits_{j \in \mathcal{N}_i^+ } \frac{{\bf h}_j^{U(l-1)}}{|\mathcal{N}_i^+ |}, \sum\limits_{k \in \mathcal{N}_i^-} \frac{{\bf h}_k^{B(l-1)}}{|\mathcal{N}_i^-|}, {\bf h}_i^{U(l-1)}\Big] \Bigg) 
\end{align}

where ${\bf W}^{B(l)},{\bf W}^{B(l)} \in \mathbb{R}^{d^{out} \times 3d^{out}}$ for $l>1$.  Note that we are utilizing the same logic here as when defining the sets $B_i(l)$ and $U_i(l)$. When gathering user $u_i$'s ``friend'' representation (i.e., ${\bf h}^{B(l)}_i)$ at layer $l$ (when $l>1$) it is based upon aggregating the ``friend'' representation at layer $(l-1)$ (i.e., ${\bf h}^{B(l-1)}_k$) for all positively linked neighbors $u_j \in \mathcal{N}^{+}_i$ while simultaneously collecting the average amongst the ``enemy'' level $(l-1)$ (i.e, ${\bf h}^{U(l-1)}_j$) information from all negatively linked neighbors $u_k \in \mathcal{N}^{-}_i$. Thus, for the case when $l=2$ we can see the ``friend'' representation is in fact gathering information from not only their direct friends (i.e., positively linked neighbors), but also (at the two hop level) friends of friends', and enemies of enemies'. Similarly, in the case of $l=2$ our hidden representation ${\bf h}^{U(l)}_i$ (i.e., user $u_i$'s ``enemy'' representation), the first layer would have gathered direct negatively linked neighbor information, but in the second layer, we are gathering from $u_i$'s friends' enemies and their enemies' friends. 

With the above discussed aggregation methods, we can now present the entire framework of SGCN. First, in Algorithm~\ref{alg:sgcn} we discuss how to obtain the embedding for each user $u_i$ in the signed network. On line 1, we set ${\bf h}^{(0)}_i$ equal to ${\bf x}_i$ for ease in defining the rest of the algorithm. Then on lines 2 through 5 we show the first layers aggregation process. Next, if the total number of layers in the SGCN is greater than one (i.e, $L>1$), then we perform the subsequent aggregations according to the defined higher level aggregation functions we designed based on balance theory. Finally, on line 14 the last step is concatenating the two hidden representations for user $u_i$, namely ${\bf h}^{B(L)}_i$ and ${\bf h}^{U(L)}_i$ together into a single low-dimensional representation.

 \begin{algorithm}[]
 \small
 	\DontPrintSemicolon
 	\KwIn{$\mathcal{G}=(\mathcal{U},\mathcal{E}^+,\mathcal{E}^-)$; an initial seed node representation \{${\bf x}_i, \forall u_i \in \mathcal{U}\}$; number of aggregation layers L; weight matrices ${\bf W}^{B(l)}$ and ${\bf W}^{U(l)}$, $\forall l \in \{1, \dots ,L\}$; non-linear function $\sigma$}
 	\KwOut{Low-dimensional representations ${\bf z}_i,  \forall u_i \in \mathcal{U}$}
    ${\bf h}^{(0)}_{i} \leftarrow {\bf x}_i, \forall u_i \in \mathcal{U}$\;

	\For{$u_i \in \mathcal{U}$}
    {
        ${\bf h}_i^{B(1)} \leftarrow \sigma \Bigg( {\bf W}^{B(1)} \Big[ \sum\limits_{j \in \mathcal{N}_i^+ } \frac{{\bf h}_j^{(0)}}{|\mathcal{N}_i^+ |}, {\bf h}_i^{(0)} \Big]  \Bigg)$\;
		${\bf h}_i^{U(1)} \leftarrow \sigma \Bigg( {\bf W}^{U(1)} \Big[ \sum\limits_{k \in \mathcal{N}_i^-} \frac{{\bf h}_k^{(0)}}{|\mathcal{N}_i^-|}, {\bf h}_i^{(0)} \Big]  \Bigg)$\;

    }
    \If{L $> 1$}
    {
    	\For{$l=2 \dots L$}
    	{
    		\For{$u_i \in \mathcal{U}$}
        	{
				${\bf h}_i^{B(l)} = \sigma \Bigg( {\bf W}^{B(l)} \Big[ \sum\limits_{j \in \mathcal{N}_i^+} \frac{{\bf h}_j^{B(l-1)}}{|\mathcal{N}_i^+|}, \sum\limits_{k \in \mathcal{N}_i^-} \frac{{\bf h}_k^{U(l-1)}}{|\mathcal{N}_i^-|}, {\bf h}_i^{B(l-1)}\Big] \Bigg)$

				${\bf h}_i^{U(l)} = \sigma \Bigg( {\bf W}^{U(l)} \Big[ \sum\limits_{j \in \mathcal{N}_i^+ } \frac{{\bf h}_j^{U(l-1)}}{|\mathcal{N}_i^+ |}, \sum\limits_{k \in \mathcal{N}_i^-} \frac{{\bf h}_k^{B(l-1)}}{|\mathcal{N}_i^-|}, {\bf h}_i^{U(l-1)}\Big] \Bigg)$ 
        	}
    	}
    }
	${\bf z}_i \leftarrow [{\bf h}^{B(L)}_i,{\bf h}^{U(L)}_i], \forall u_i \in \mathcal{U}$\;
  \caption{Signed GCN Embedding Generation.}
 \label{alg:sgcn}
 \end{algorithm}

Next we design an objective function to learn the parameters of SGCN. The objective function for SGCN is based upon two components, both of which are based on the goal that we would like the representations to be able to understand the relationships between pairs of users in the signed network's embedded space. The first term incorporates an additional layer for performing a weighted multinomial logistic regression (MLG) classifier. Here we wish to classify whether a pair of node embeddings are from users with a positive, negative, or no link between them. More specifically, we construct a mini-batch of users and then a set $\mathcal{M}$, which contains triplets of the form $(u_i,u_j,s)$ which denotes the pair of users ($u_i$,$u_j$) along with $s \in \{+,-,?\}$ for denoting whether there was a positive, negative, or no link between the pair of users. For input into the classifier, we use the final embeddings for users $u_i$ and $u_j$ concatenated together (i.e,. [${\bf z}_i, {\bf z}_j$]). We use $\omega_s$ to denote the weight associated with class $s$. We introduce a second term that is founded on extended structural balance theory. This term is controlled by $\lambda$ to balance the contribution towards the overall objective. The goal of this second term is to  have positively linked users closer in the embedded space than the no link pairs, and the no link paired users should be closer than users having a negative link between them. The overall objective is formalized in the following:

\begin{align}\label{eq:loss}
&\mathcal{L}(\theta^{W},\theta^{MLG}) = \nonumber \\
&- \frac{1}{\mathcal{M}}  \sum\limits_{(u_i,u_j,s) \in \mathcal{M}} \omega_s \log \frac{\exp{([{\bf z}_i,{\bf z}_j]\theta^{MLG}_s)}}{\sum\limits_{q \in \{+,-,?\}} \exp{([{\bf z}_i,{\bf z}_j]\theta^{MLG}_q )}} \nonumber \\
&+ \lambda \Bigg[ \frac{1}{|\mathcal{M}_{(+,?)}|} \sum\limits_{\substack{(u_i,u_j,u_k) \\ \in \mathcal{M}_{(+,?)}}} \max\Big(0,(|| {\bf z}_i - {\bf z}_j||^2_2 - || {\bf z}_i - {\bf z}_k||^2_2) \Big) \nonumber \\
&  + \frac{1}{|\mathcal{M}_{(-,?)}|} \sum\limits_{\substack{(u_i,u_j,u_k) \\ \in \mathcal{M}_{(-,?)}}} \max\Big(0,(|| {\bf z}_i - {\bf z}_k||^2_2 - || {\bf z}_i - {\bf z}_j||^2_2) \Big) \Bigg] \nonumber \\
&+ Reg(\theta^{W},\theta^{MLG})
\end{align}

$\theta^{W}$ represents the weight matrices used in the layers of our SGCN, $\theta^{MLG}$ denotes the parameters of the MLG classifier, $\omega_s$ is used for the weight associated with the class $s$ (with $s \in \{+,-,?\}$ for the positive, negative, and no link classes), $\mathcal{M}_{(+,?)}$ and $\mathcal{M}_{(-,?)}$ are the sets for the pairs of positive and negatively linked users, respectively, where for every linked pair $(u_i,u_j)$ we further sample another user $u_k$ randomly (and different in each epoch) that has no link to $u_i$. The term $Reg(\theta^{W},\theta^{MLG})$ we use for regularization on the parameters of our model. For updating the parameters, we utilize the same SGD style updating as presented in~\cite{Hamilton-etal2017}, since it has been show to effectively update the parameters of a GCN using a mini-batch setting (as compared to previous work such as in~\cite{Kipf-Welling2016} that performed batch gradient descent).  

\section{Experiments}\label{sec:experiments}
In this section, we experimentally evaluate the effectiveness of the proposed signed graph convolutional network (SGCN) in learning node representations. We seek to answer the following questions: (1) Is SGCN capable of learning meaningful low-dimensional representations? and (2) Does the introduction of balance theory into the aggregation process along with longer path information provide a performance increase in learning the node embeddings?

To address the first question, we conduct experiments to measure the learned embedding quality by performing the most fundamental signed network analysis task, namely link sign prediction~\cite{Leskovec-etal2010}, and compare against the signed network embedding state-of-the-art baseline methods. To answer the second question, we investigate variants of our framework that do not exploit the longer paths (i.e., only performing a single aggregation step) or that do not make use of balance theory (i.e., the fundamental signed social network theory). 

\subsection{Experimental Settings}
In this subsection, we begin by introducing our datasets, the link sign prediction problem, and the metrics used for evaluation.

For our study of learning representations using signed graph convolutional networks, we conduct our experiments on four real-world signed network datasets, i.e., Bitcoin-Alpha\footnote{http://www.btcalpha.com}, Bitcoin-OTC\footnote{http://www.bitcoin-otc.com}, Slashdot\footnote{http://www.slashdot.com} and Epinions\footnote{http://www.epinions.com}. We obtained the Bitcoin-Alpha and Bitcoin-OTC datasets from~\cite{Kumar-etal2016}. These two datasets are both coming from sites that focus on having an open market where users can buy and sell things using Bitcoins. Due to the fact Bitcoin accounts are anonymous, the users of the two sites have started online trust networks for their safety. This can allow users to positively (or negatively) rate others they trust (or distrust) which can help alleviate the problem of scammers. Our third
dataset, Slashdot, is a technology news site where users can create friend (positive) and foe (negative) links between each other. The final dataset we collected is from Epinions, which is a popular product review website. This site allows their users to categorize other users into those they trust or distrust and are similarly represented as positive and negative links, respectively. We note that for each of these datasets we perform our experiments on the undirected signed networks and have further filtered out users randomly from the two larger networks (Slashdot and Epinions) that had very few links. We summarize these datasets in Table~\ref{tab:datasets} with some basic statistics.

The problem of predicting the signs of links~\cite{Leskovec-etal2010} is that given a set of existing links in the signed network that had been held out of the training set, we wish to predict their signs being positive or negative between those pairs of users. Thus, a binary classifier is used to predict the sign based on a set of input features from the pair of users (more specifically we employ a logistic regression model). In our case we concatenate the final embeddings of the two users together as the set of features. The model is trained using the labeled edges from the training data. For evaluation, since the positive and negative links are unbalanced (i.e., there are many more positive links than negative links), we utilize both F1 and Area Under the receiver operating characteristic Curve (AUC). We note that higher F1 and AUC both mean better performance. For each dataset, we randomly choose 20\% of the data as test, and the remaining 80\% as training. Note that we used a grid search along with cross validation on the training data to tune the hyperparameters of our model.

\subsection{Performance Comparison}
Here we present some existing state-of-the-art signed network embeddings methods such that we can study the effectiveness of our signed GCN (SGCN) in learning node representations in signed networks. For succinctness we do not include unsigned methods since previous signed network embedding work has shown their superiority over the non-dedicated efforts towards signed network embeddings. The baselines are as follows:

\begin{itemize}
\item Signed Spectral Embedding (SSE)\cite{Kunegis-etal2010}: A spectral clustering algorithm based on the proposed signed version of the Laplacian matrix. We utilize the top-$d^{out}$ eigenvectors corresponding to the smallest eigenvalues as the embedding vectors for each node. 
\item SiNE\cite{Wang-etal2017}: This method is a deep learning framework that utilized extended structural balance theory.
\item SIDE\cite{Kim-etal2018}:
A random walk based method, utilizing balance theory, is used to obtain indirect connections for a likelihood formulation.
\end{itemize}

Furthermore, we propose to evaluate the following two variants of our model:
\begin{itemize}
\item SGCN-1: This method only makes use of the first single aggregation layer and therefore only separates the positive from the negative links (i.e, does not yet make use of balance theory and our defined balanced paths). 
\item SGCN-1+: This method similar to SGCN-1 does not make use of balance theory, instead it performs the na\"ive aggregation of the first layer, but twice. In other words, the final representation for each user is based on propagating information along the positive links twice, and the negative links twice, separately.
\end{itemize}

Some final notes are the following: 1) in our experiments we do not have node attributes, therefore instead we use the final embedding of the SSE model as the input feature matrix (i.e., ${X})$  to all our SGCN variants; 2) for all embedding methods we fixed the final low-dimensional representation to be 64; 3) We used the authors released code for SiNE\footnote{http://www.public.asu.edu/~swang187/codes/SiNE.zip} and use their suggested hyperparameters~\cite{Wang-etal2017} for our experiments; 4) For SIDE, we use the authors implementation\footnote{https://datalab.snu.ac.kr/side/resources/side.zip} and the suggested hyperparameter settings from~\cite{Kim-etal2018}, but for the unsuggested parameters we used a grid search around their code's default settings; and 5) for our models we set $\lambda = 5$ and the ``friend'' and ``enemy'' hidden representations were each set to 32, such that the final embeddings were of size 64.

\begin{table}	
	\begin{center}
		\caption{Statistics of four signed social networks.}
		\label{tab:datasets}
		\begin{tabular}{c|c|c|c}
			\hline
			Network & \# Users & \begin{tabular}{c}\# Positive\\ Links\end{tabular} & \begin{tabular}{c}\# Negative\\ Links\end{tabular} \\ \hline 
			Bitcoin-Alpha  & 3,784 & 12,729  & 1,416     \\	
			Bitcoin-OTC & 5,901 & 18,390  & 3,132     \\	
			Slashdot  & 33,586 & 295,201  & 100,802     \\	
			Epinions  & 16,992 & 276,309  & 50,918 \\ \hline
		\end{tabular}
	\end{center}
	\vspace{-2ex}
\end{table}

\begin{table}
	\caption{Link Sign Prediction Results with AUC.}\label{tab:link_prediction_results_auc}
	\centering
	\begin{tabular}{|c|c|c|c|c|} \hline
		\begin{tabular}{@{}c@{}}Embedding \\ Method\end{tabular}& Bitcoin-Alpha&Bitcoin-OTC&Slashdot&Epinions\\ \hline
	    SSE		&	0.764		&	0.803		&	0.769		&	0.822 		\\ \hline
		SiNE	&	0.778		&	0.814		&	0.792		&	0.849 		\\ \hline
		SIDE	&	0.630		&	0.618		&	0.547			&	0.571		\\ \hline
		SGCN-1	&	0.780		&	0.818		&	0.784		&	0.663 		\\ \hline
		SGCN-1+	&   0.785   	&   0.817  		&   {\bf 0.804}	&   0.722 		\\ \hline
		SGCN-2	&	{\bf 0.796}	&	{\bf 0.823}	&	{\bf 0.804}		&	{\bf 0.864} \\ \hline
	\end{tabular}
\end{table}

\begin{table}[h]
	\caption{Link Sign Prediction Results with F1.}\label{tab:link_prediction_results_f1}
	\centering
	\begin{tabular}{|c|c|c|c|c|} \hline
		\begin{tabular}{@{}c@{}}Embedding \\ Method\end{tabular}& Bitcoin-Alpha&Bitcoin-OTC&Slashdot&Epinions\\ \hline
		SSE		&	0.898		&	0.923		&	0.820		&	0.901 		\\ \hline
		SiNE	&	0.888		&	0.878		&	0.854		&	0.914 		\\ \hline
		SIDE	&	0.738		&	0.750		&	0.646			&	0.711		\\ \hline
		SGCN-1	&	0.910		&	0.918		&	0.853		&	0.851 		\\ \hline
		SGCN-1+ &   0.912   	&   0.923  		&   {\bf 0.865}	&   0.893 		\\ \hline
		SGCN-2	&	{\bf 0.917}	&	{\bf 0.925}	&	0.864		&	{\bf 0.933}	\\ \hline
	\end{tabular}
\end{table}

\subsection{Comparison Results}
 The comparison results in terms of AUC and F1 are demonstrated in Tables~\ref{tab:link_prediction_results_auc} and~\ref{tab:link_prediction_results_f1}, respectively.  For the tables, we make the following observations: 
\begin{itemize}
\item SGCN-1 with only one step aggregation from positive and negative links obtains comparable performance with the best performance from the baselines. This observation suggests that it is necessary to separate positive and negative links. 
\item SGCN-1+ outperforms SGCN-1. The results indicate that propagating multiple steps during the aggregation can help improve the performance.
\item Most of the time, SGCN-2 outperforms SGCN-1 and SGCN-1+. Aggregation following the longer  balance and unbalanced paths can boost the performance. 
\end{itemize}

\subsection{Parameter Analysis}
The proposed signed GCN has one major hyperparameter, $\lambda$ (besides the number of layers and the aggregation types which we have already investigated with our variants of SGCN). The parameter $\lambda$ is used to control the balance between the two terms in our objective function as given in Eq.~(\ref{eq:loss}). More specifically, the first term introduced the multinomial logistic regression term in an attempt to guide the learned node embedding to be separable such that pairs of user that have positive, negative, and no link can be positioned such that the classifier can distinguish their relationship. The second term we utilized for discovering node embeddings that adhere to extended structural balance theory~\cite{Cygan-etal2015}. With its contribution controlled by $\lambda$, this term forced pairs of users that have positive links to be closer in the low-dimensional embedding space than to other users they had no link with, and further also sought to have users with negative links pulled further apart by wanting no linked pairs closer together than the negative pairs. 

\begin{figure}
\begin{center}
\subfigure[SGCN-2 (F1) ]{\label{fig:f1}\includegraphics[scale=0.2]{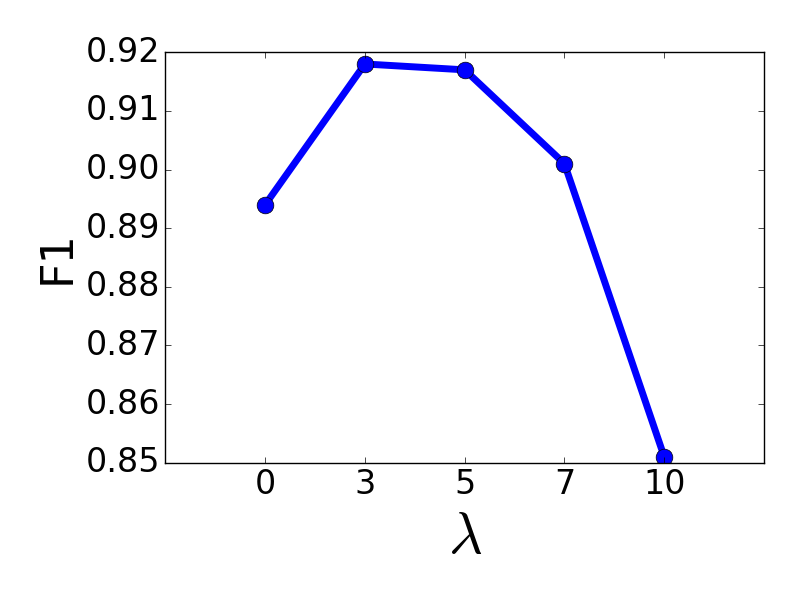}}
\subfigure[SGCN-2 (AUC)]{\label{fig:auc}\includegraphics[scale=0.2]{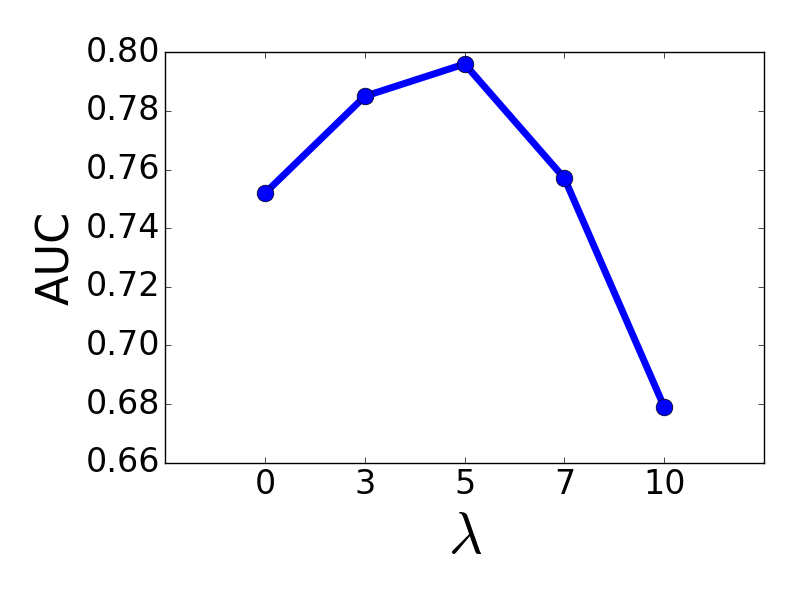}}
\end{center}
\caption{Parameter Sensitivity when varying the parameter $\lambda$ on the Bitcoin-Alpha dataset.}
\end{figure}

In Figures~\ref{fig:f1} and~\ref{fig:auc} we report the results when varying $\lambda$ for one of the signed network datasets, namely Bitcoin-Alpha. We do not show results of other settings since we can have similar observations.  As we can see from these two figures $\lambda = 5$ seems to be a good balance between the AUC and F1 performance. The second observation is that when setting $\lambda$ equal to zero we have a drastic decrease in performance. Note that we saw similar results across all datasets in that the contribution of the second term, which is based on balance theory, was able to provide an improvement.

\section{Related Work}\label{sec:relatedwork}
In this section, we present and discuss related work on signed networks (with a primary focus on signed network embedding) and on the recent development of graph convolutional networks. 

\subsection{Signed Network Embedding}
With the prevalence of online social media, signed network analysis has become ubiquitous and thus attracted an increased attention~\cite{Tang-etal2016}. The roots of signed network analysis were developed in social psychology~\cite{Heider-1946,Cartwright-Haray1956} and since then many applications and measurements such as link sign prediction~\cite{Leskovec-etal2010,Chiang-etal2011} and signed centrality~\cite{Kunegis-etal2009,Shahriari-Jalili2014}, respectively, have been the primary focus in this domain. However, some directions such as network modeling and network embedding have yet to be fully explored for signed networks. 

The goal of network embedding (or representation learning) is to learn a low-dimensional representation for all nodes in a given network. These vector representations can thereafter be used in many tasks such as node classification, link prediction, and even provide a visualization of the networks. 
Most of the attention in this domain has focused on learning representations for unsigned networks~\cite{Grover-Leskovec2016,Tang-etal2015,Perozzi-etal2014,Hamilton-etal2017}, which do not take negative links into consideration. However, negative links have been shown to provide added value over only using positive links~\cite{Leskovec-etal2010,Ma-etal2009}. In~\cite{Kunegis-etal2010} they extended spectral analysis for signed networks and utilized the top-k eigenvectors as the node embeddings for tasks such as signed link prediction and visualization. Later, a matrix factorization approach was used for extracting latent vectors for each user in a signed network~\cite{Chiang-etal2012}. The problem was then not investigated much until SiNE~\cite{Wang-etal2017}, which developed a deep learning framework that utilized extended structural balance theory\cite{Cygan-etal2015}. Around the same time SNE~\cite{Yuan-etal2017} was developed utilizing a log-bilinear model and sampling from random walks. 

Thereafter, works were introduced to handle the attributed~\cite{Wang-etal2017a} and directed signed network settings~\cite{Kim-etal2018}. In the former, node attributes are utilized in the SNEA framework~\cite{Wang-etal2017a} to obtain a better performance on tasks such as node clustering by enforcing a constraint that users with similar attributes should have similar representations (along with using extended structural balance theory on the signed links). The later, SIDE~\cite{Kim-etal2018}, which was proposed most recently, provides a linearly scalable method (in relation to the number of nodes) that leverages balance theory along with random walks to obtain longer indirect connections for their likelihood formulation. Recently a signed heterogeneous network embedding algorithm, SHINE~\cite{Wang-etal2018}, had been developed, but in their work the social network itself is unsigned. They include signed information by extracting the sentiment associated with user posted text and further include user profile attribute information; thus leaving their method not applicable to the traditional signed network embedding problem. 

\subsection{Graph Convolution Networks}
Recently, there has been an increased interest in utilizing convolutional neural networks on graph data~\cite{Duvenaud-etal2015,Defferrard-etal2016,Kipf-Welling2016,Niepert-etal2016,Bruna-etal2014,Hamilton-etal2017}. The early works on using convolutional neural networks for graphs focused on employing them on the entire network (and hence do not scale to larger networks) and/or designed for learning representations for the whole entire network for graph classification~\cite{Niepert-etal2016,Bruna-etal2014,Duvenaud-etal2015,Defferrard-etal2016} (as compared to learning node representations). The original GCN algorithm~\cite{Kipf-Welling2016}, that was designed with the focus to learn representations at the node level, was introduced in a semi-supervised learning framework for mapping node features (i.e., attributes) to node classes (i.e., the node classification problem). More recently, GraphSAGE~\cite{Hamilton-etal2017} extended the ideas presented in~\cite{Kipf-Welling2016} by first providing an inductive setting. This allows for learning node representations of previously unseen data (i.e., nodes that arrive in the network after the training process) by mapping the node features through the GCN to obtain a low-dimensional representation and furthermore the class label for the given node. They also introduced numerous aggregation functions and an efficient strategy to utilize stochastic gradient descent (SGD) rather than requiring the entire network in memory per update as required by the previous work.  However, all the above mentioned GCNs have only been defined to handle unsigned networks.

\section{Conclusion}\label{sec:conclusion}
In conclusion, recently there has been a growing interest in the utilization of graph convolutional networks (GCNs) due to the fact they have been shown to provide great improvements in many tasks such a network embedding. Simultaneously, signed social networks have become ubiquitous with the growing popularity of online social media. However, most previous work in GCNs has been on unsigned networks, which are inherently unable to handle the complexities and challenges associated with the inclusion of the negative links found in signed networks. Therefore dedicated efforts are required if we wish to harness the power of GCNs to perform signed network related tasks, such as signed network embedding. 

Although it was clear GCNs would need redesigned for signed networks, there were still tremendous challenges to overcome. More specifically, they were on how to handle the negative links and furthermore how to combine the positive and negative links together into a single coherent model. The key to overcoming these challenges we discovered lie in the roots of signed network analysis. We proposed the use of the fundamental social psychology theory designed to provide insights into how positive and negative links interact in the complex signed networks; more specifically, we utilized balance theory. This allowed us to bridge the gap between the recent advances in unsigned GCNs and the domain of signed network analysis. Using our constructed signed graph convolutional network, we performed empirical evaluations through experiments on four real-world signed networks. Comparing against the state-of-the-art signed network embedding algorithms, we have shown the superiority of the SGCNs when performing the classical link sign prediction task.

For our future work, we first plan to further investigate the usage of SGCNs for other tasks in signed networks beyond node embeddings, such as specific efforts towards node classification. Thereafter, we will focus on using other deep learning architectures for constructing a deep generative network model for signed networks.

\bibliographystyle{IEEEtran}
\bibliography{reference/sgcn.bib}

\end{document}